\documentclass[aps,twocolumn,noshowkeys]{revtex4-1}
\usepackage{amsmath}
\usepackage{graphicx}
\usepackage{bm}
\usepackage{hyperref}
\usepackage{xcolor}

\newcommand{\Eref}[1]{Eq.~(\ref{#1})}
\newcommand{\tref}[1]{Table~\ref{#1}}

\newcommand{\WaxeN}{W_{\textrm{ax}}^{(eN)}}

\begin{document}
\title{Electronic structure of ytterbium monohydroxide molecule to search for axionlike particles}

\begin{abstract}
Recently, the YbOH molecule has been suggested as a candidate to search for the electron electric dipole moment (eEDM), which violates spatial parity ($P$) and time-reversal ($T$) symmetries [I.~Kozyryev and N.~R.~Hutzler, Phys. Rev.  Lett. \textbf{119}, 133002 (2017)]. In the present paper, we show that the same system can be used to measure coupling constants of the interaction of electrons and nucleus mediated by axionlike particles. The electron-nucleus interaction produced by the axion exchange can contribute to a $T,P$-violating EDM of the whole molecular system. We express the corresponding $T,P$-violating energy shift produced by this effect in terms of the axion mass and product of the axion-nucleus and axion-electron coupling constants.
\end{abstract}

\author{D.E. Maison$^{1,2,*}$, 
V.V.\ Flambaum$^{3,4}$,
N.R.\ Hutzler$^{5}$,
L.V.\ Skripnikov$^{1,2}$}
\affiliation{$^{1}$Petersburg Nuclear Physics Institute named by B.P.\ Konstantinov of National Research Center ``Kurchatov Institute'' (NRC ``Kurchatov Institute'' - PNPI), 1 Orlova roscha mcr., Gatchina, 188300 Leningrad region, Russia}
\affiliation{$^{2}$Saint Petersburg State University, 7/9 Universitetskaya nab., St. Petersburg, 199034 Russia}
\homepage{http://www.qchem.pnpi.spb.ru}

\email{daniel.majson@gmail.com} \email{maison\_de@pnpi.nrcki.ru}

\affiliation{$^{3}$School of Physics, The University of New South Wales, Sydney NSW 2052, Australia}
\affiliation{$^{4}$Johannes Gutenberg-Universit\"at Mainz, 55099 Mainz, Germany}
\affiliation{$^{5}$Division of Physics, Mathematics, and Astronomy, California Institute of Technology, Pasadena, CA 91125, United States}

\maketitle

\section{Introduction}
Verification of the standard model (SM) and its extensions is of key importance for modern theoretical physics and cosmology~\cite{Safronova:18}. Despite numerous experimental confirmations of the SM, several observed phenomena cannot be explained in its framework. Among them are the strong $CP$-problem of quantum chromodynamics (QCD), the unknown nature of dark matter and dark energy and the problem of baryogenesis~\cite{Sakharov1967}.

A possible solution of the strong $CP$-problem was suggested by Peccei and Quinn \cite{peccei1977cp} via modification of the QCD Lagrangian. Weinberg and Wilczek also independently noticed that the spontaneous Peccei-Quinn $U_{\textrm{PQ}}(1)$ symmetry breaking demands the existence of the pseudoscalar Goldstone boson, afterwards called the axion \cite{wilczek1978problem,weinberg1978new}. Later, it has been realized that the axion is a suitable candidate to be the dark matter component \cite{preskill1983cosmology,abbott1983cosmological,dine1983not}. Therefore, it also solves another aforementioned problem of the SM. In the  QCD axion case, there are relations connecting axion mass and axion-fermion interaction strength. In a more general case of axionlike particles, no such connection is assumed.  Numerous experimental investigations led to strong constraints on the  axionlike particle properties \cite{terrano2015short,crescini2017improved,rong2017searching} (see also Fig. 2 in Ref.~\cite{stadnik:2018}). For brevity, axionlike particles are often called axions, without assuming the QCD axion properties.

It is known that the search for $T,P$-violating effects, such as the electron electric dipole moment (eEDM), in the low-energy regime can be successfully performed with paramagnetic heavy atoms and small molecules containing such atoms~\cite{Safronova:18,Ginges:04,KL95}. One of the benefits of such molecules is the existence of closely lying opposite parity levels enhancing eEDM effect \cite{Sushkov:78}. For instance, the strongest current constraint on the eEDM value $d_e$ is obtained in an experiment using a thorium monoxide molecular beam \cite{ACME:18}. Other molecular experiments~\cite{Cornell:2017, Hudson:11a} also surpass the sensitivity of atomic EDM experiments~\cite{Regan:02}. 
The constraints on the axion-fermion interaction parameters can also be obtained in  these experiments~\cite{stadnik:2018,flambaum2018resonant}.

It was recently suggested to perform experiments to search for the electron electric dipole moment using linear triatomic molecules containing heavy atoms \cite{kozyryev2017precision,isaev2016polyatomic,isaev2017laser}. These molecules have a very small energy gap between opposite parity levels due to $l$-doubling effect~\cite{kozyryev2017precision,Hutzler:2020}. This feature makes it possible to polarize them by the relatively weak electric field. Also, these molecules can be cooled and slowed by the laser-cooling technique to extremely low temperatures \cite{augenbraun2020laser}. This is the way to increase the coherence time and, thus, to improve the sensitivity of the experiment as the uncertainty of the measured energy characteristic is inversely proportional to the coherence time. For this reason, the ytterbium monohydroxide (YbOH) molecule has been intensively considered for the $T,P$-violating effect search by several theoretical groups \cite{Maison:2019b,denis2020enhanced,denis2019enhancement,prasannaa2019enhanced,gaul2020ab}. 

In Ref.~\cite{stadnik:2018} the first estimations of the axion-induced interactions in several diatomic molecules have been performed. These estimations have been based on atomic calculations and scaling. In the present paper, we introduce an explicit molecular approach to study the axion-mediated $T,P$-violating interaction based on the relativistic Fock-space coupled cluster theory. The dependence of the corresponding energy shift on the axionlike particle mass is considered in the YbOH molecule. It is shown that the expected sensitivity~\cite{kozyryev2017precision} of the experiment on this molecule may enable one to get updated restrictions for axion-fermion interaction strength.

\section{Theory}

The coupling of an axion field, denoted by $a$, with the SM fermions $\psi$ can be written in the following form: 
\begin{equation}
\label{Lagr}
\mathcal{L}_\textrm{int} = a \sum_\psi \bar{\psi} \left(g_\psi^s + ig_\psi^p \gamma_5 \right) \psi \, .
\end{equation}
The coupling constants ~$g_\psi^s$ and ~$g_\psi^p$ characterize the scalar and  pseudoscalar interactions in the Lagrangian (\ref{Lagr}). This mixed scalar and pseudoscalar interaction leads to $T,P$-violating effects.

The electron pseudoscalar and nucleon scalar interactions in Eq.~(\ref{Lagr}) with the intermediate boson $a$ of the mass  $m_a$ lead to the $T,P$-violating Yukawa-type interaction potential~\cite{moody1984new,stadnik:2018}:
\begin{equation}
    \label{Yukawa}
    V_{\textrm{eN}}(\boldsymbol{r}) = +i \frac{g_N^s g_e^p}{4 \pi } \frac{e^{-m_a  |\boldsymbol{r}-\boldsymbol{R}|}}{|\boldsymbol{r}-\boldsymbol{R}|} \gamma_0 \gamma_5.
\end{equation}
Here $\boldsymbol{r}$ and $\boldsymbol{R}$ are positions of the electron and nucleon under consideration, respectively; $\gamma$ matrices are Dirac matrices defined according to Ref. \cite{Khriplovich:91} and refer to the electron; $g_N^s$ and $g_e^p$ are the coupling constants of the axionlike particle with the nucleon $N$ (either proton or neutron) and the electron, respectively. 
This interaction has a similar form as the $T,P$-violating contact nucleus-electron scalar-pseudoscalar interaction \cite{Ginges:04}:
\begin{eqnarray}
  H_{T,P}=+i\frac{G_F}{\sqrt{2}}Zk_{T,P} n(\textbf{r}) \gamma_0 \gamma_5,
 \label{Htp}
\end{eqnarray}
where $G_F$ is the Fermi-coupling constant, $Z$ is the nuclear charge, $k_{T,P}$ is the coupling constant and $n(\textbf{r})$ is the nuclear density normalized to unity.

The nonrelativistic limit of \Eref{Yukawa} leads to the interaction which is proportional to the scalar product $(\nabla a \cdot \mathbf{s})$, where $\mathbf{s}$ is the electron spin. Thus, this interaction can be described as the interaction of the axion with the electron spin. The opposite case of the interaction of the axion with the nucleon spin, described by the product $g_e^s g_N^p$, is considered in, e.g. Refs. \cite{moody1984new, dzuba2018new}.

Inclusion of the interaction~(\ref{Yukawa}) into the electronic Hamiltonian leads to $T,P$-violating energy shifts of electronic states in a manner analogous to the shifts created by the nucleus-electron scalar-pseudoscalar interaction. This shift can be characterized by the molecular constant $\WaxeN$, which depends on the axionlike particle mass $m_a$:
\begin{equation}
    \WaxeN(m_a) = 
    \frac{1}{\Omega}
    \langle \Psi | \sum\limits_{i=1}^{N_\textrm{e}} \sum\limits_{N} 
    \frac{1}{g_N^s g_e^p}
    V_{\textrm{eN}}(\boldsymbol{r}_i)
    | \Psi \rangle.
\end{equation}
Here $\Psi$ is the electronic wave function, $N_\textrm{e}$ is number of electrons, index $i$ runs over all the electrons in the molecule,  the inner sum is over all the nucleons in the heavy nucleus (the analogous interaction with light nuclei is omitted in the present paper) and $\Omega$ is the projection of electronic momentum on the molecular axis. For the state considered here, $\Omega = 1/2$. The characteristic $T,P$-violating energy shift of the electronic level can be expressed as
\begin{equation}
\label{Etp}
    \delta E = \bar{g}_{N}^s g_e^p  \Omega \WaxeN(m_a),
\end{equation}
where $\bar{g}_{N}^s$ is the corresponding constant averaged over Yb nucleus nucleons: $\bar{g}_{N}^s = (Zg_p^s+N_n g_n^s)/A$; $g_p^s$ and $g_n^s$ are scalar axion-proton and axion-neutron coupling constants, $Z$ is the charge of the heavy nucleus, $N_n$ is its neutron number and $A=Z+N_n$.
The value of $\WaxeN$ is required for interpretation of the experimental data in terms of the product of interaction constants. This molecular constant is the analog of the effective electric field and the molecular constant $W_{T,P}$, which are used to interpret experiments to search for $T,P$-violating atomic and molecular EDMs in terms of the electron electric dipole moment and the scalar-pseudoscalar nucleus-electron interaction constant (see, e.g. Refs.~\cite{Titov:06amin,Skripnikov:16b,Skripnikov:14c,Skripnikov:17c,Fleig:17,Skripnikov:15a,denis2019enhancement,prasannaa2019enhanced}).
Note, however, that in the present case $\WaxeN$ depends on the axion mass. The typical radius of the interaction \Eref{Yukawa} can be estimated as $R_{\textrm{Yu}}(m_a) \simeq \frac{3730}{(m_a/\textrm{eV})} a_B$, 
where $a_B$ is the Bohr radius. 
As it is shown below, in the limiting case 
of large $R_{\textrm{Yu}}$, the $\WaxeN$ constant is almost independent of $m_a$ and in the opposite limiting case of high-mass axionlike particles, the factorization of $\WaxeN$ is possible.

In the present paper, we calculate molecular constant $\WaxeN(m_a)$ for the YbOH molecule over a wide range of $m_a$ values. The molecule is considered in its ground $^2\Sigma_{1/2}$ electronic state. 
The equilibrium geometry parameters of the molecule are $R(\textrm{Yb-O}) = 2.037 \AA$ and $R(\textrm{O-H}) = 0.951 \AA$; the molecule is linear \cite{brutti2005mass, nakhate2019pure}. 
As it was shown in Refs. \cite{gaul2020ab,zakharova:2020}, electronic properties of the molecule in its ground state are close to the ones in the metastable bending mode, which are of the main interest.

In the electronic structure calculations, we have used one-particle molecular bispinors obtained within the Dirac-Hartree-Fock approach using the Gaussian-type basis sets. In order to estimate the basis set size dependence of the $\WaxeN$ parameter, the calculations were performed within four basis sets, which are described in Table~\ref{Basises}. These basis sets are ordered by quality, i.e., the basD is the best considered basis set.
For Yb we have used Dyall's family of all-electron basis sets.

\begin{table}[h]
    \caption{Notation and composition of the basis sets used.}
    \label{Basises}
    \begin{tabular}{lll}
    \hline
    \hline
         Basis set & 
         Basis on  & 
         Basis on  \\
         notation & 
         Yb \cite{gomes:2010}$^*$ &  
         O and H \cite{Kendall:92,Dunning:89,de2001parallel}$^*$ \\
         \hline
         \hline
         basA & AE2Z & aug-cc-pVDZ-DK \\ 
           & [24$s$,19$p$,13$d$,8$f$,2$g$] & [10$s$,5$p$,2$d$] and [5$s$,2$p$] \\
         \hline
         basB & AE3Z & aug-cc-pVDZ-DK \\ 
          & [30$s$,24$p$,16$d$,11$f$,4$g$,2$h$] &
          [10$s$,5$p$,2$d$] and [5$s$,2$p$] \\
         \hline
         basC & AE3Z & aug-cc-pVTZ-DK\\
          & 
          [30$s$,24$p$,16$d$,11$f$,4$g$,2$h$]
         & [11$s$,6$p$,3$d$,2$f$] and [6$s$,3$p$,2$d$] \\
         \hline
         basD & AE4Z & aug-cc-pVTZ-DK\\
         & [35$s$,30$p$,19$d$,14$f$,8$g$,5$h$,2$i$] & 
         [11$s$,6$p$,3$d$,2$f$] and [6$s$,3$p$,2$d$] \\
    \hline
    \hline
    \end{tabular}
    $^*$All the basis sets were taken in the uncontracted form.
\end{table}

Electronic correlation effects have been taken into account using the relativistic Fock-space coupled cluster approach with single and double cluster amplitudes (FS-CCSD) \cite{Visscher:01}.
Fock-space sector (0,0) corresponds to the YbOH$^+$ cation in its ground electronic state, and the open-shell electronic calculations were performed in sector (0,1). All electrons of YbOH have been included in correlation calculations. In Refs.~\cite{Skripnikov:17a,Skripnikov:15a} it has been shown that high energy cutoff is important to ensure including functions that describe spin-polarization and correlation effects for core electrons. In the present paper, all virtual orbitals have been included in correlation treatment. The effect of the Gaunt interaction of the electrons was estimated within the Dirac-Hartree-Fock-Gaunt approach using the basD basis set. Its relative contribution reaches the maximal value of $-$2.2\% for $m_a = 10^4 \ \textrm{eV}$ and does not exceed 1\% by absolute value for other presented $m_a$. It was shown in Ref. \cite{Maison:20a} that electronic correlation can affect the Gaunt contribution to the properties of triatomic molecules, but the absolute value contribution is not very significant. We do not include the Gaunt contribution to the values in Table~\ref{TResult1}.
In order to calculate the value of the $\WaxeN$ constant, the finite field approach has been used.

Both the Dirac-Hartree-Fock and the Fock-space calculations were performed using the local version of the {\sc dirac15} code \cite{DIRAC15}. The code to calculate matrix elements of the electron-nucleus interaction \Eref{Yukawa} was developed in the present paper.

\section{Results and discussion}

Tables \ref{TResult0} and \ref{TResult1} give the calculated dependence of the constant $\WaxeN$ on the mass of the axionlike particle $m_a$ using different basis sets and methods. The final values are given in the last column of Table~\ref{TResult1}. 
The uncertainty of the final values arises mainly from higher-order correlations and can be estimated to be less than 10\% \cite{Maison:2019b}.
\begin{table}[h]
    \caption{ 
The values of the $\WaxeN$ constant for the ground electronic state of YbOH (in units of $m_e c / \hbar$) for various $m_a$ using the Dirac-Hartree-Fock method and different basis sets.}
\label{TResult0}    
    \begin{tabular}{lllll}
        \hline
        \hline
        $m_a$, eV &$\ $ basA $ \ $ & basB $ \ $& basC $ \ $ & basD\\
        \hline
        $10$  & $+ 2.84 \times 10^{-5}$ 
        & $+2,82 \times 10^{-5}$ 
        & $+2.82 \times 10^{-5}$ 
        & $+2.82 \times 10^{-5}$
        \\
        $10^2$  & $+2.82 \times 10^{-5}$ 
        & $+2.82 \times 10^{-5}$ 
        & $+2.82 \times 10^{-5}$ 
        & $+2.82 \times 10^{-5}$ 
        \\
        $10^3$  & $ +2.34 \times 10^{-5}$ 
        & $ +2.32 \times 10^{-5}$ 
        & $ +2.32 \times 10^{-5}$ 
        & $ +2.32 \times 10^{-5}$
        \\
        $10^4$  & $ +3.42 \times 10^{-6}$ 
        & $ +3.40 \times 10^{-6}$ 
        & $ +3.38 \times 10^{-6}$ 
        & $ +3.38 \times 10^{-6}$
        \\
        $10^5$  & $-1.28 \times 10^{-5}$ 
        & $ -1.27 \times 10^{-5}$ 
        & $ -1.27 \times 10^{-5}$ 
        & $ -1.27 \times 10^{-5}$
        \\
        $10^6$  & $-9.62 \times 10^{-6}$ 
        & $ -9.54 \times 10^{-6}$ 
        & $ -9.54 \times 10^{-6}$ 
        & $ -9.52 \times 10^{-6}$
        \\
        $10^7$  & $-4.06 \times 10^{-7}$ 
        & $ -4.06 \times 10^{-7}$ 
        & $ -4.06 \times 10^{-7}$ 
        & $ -4.04 \times 10^{-7}$
        \\
        $10^8$  & $ -6.78 \times 10^{-9}$ 
        & $ -7.24 \times 10^{-9}$ 
        & $ -7.24 \times 10^{-9}$
        & $ -7.32 \times 10^{-9}$
        \\
        $10^9$  & $-7.16 \times 10^{-11}$ 
        & $ -7.84 \times 10^{-11}$ 
        & $ -7.84 \times 10^{-11}$
        & $ -8.06 \times 10^{-11}$
        \\
        $10^{10}$  & $ -7.18 \times 10^{-13}$ 
        & $ -7.88 \times 10^{-13}$ 
        & $ -7.88 \times 10^{-13}$ 
        & $ -8.06 \times 10^{-13}$ 
        \\
    \hline
    \end{tabular}
\end{table}

\begin{table}[h]
\caption{
The values of the $\WaxeN$ constant for the ground electronic state of YbOH (in units of $m_e c / \hbar$) for various $m_a$ using the relativistic FS-CCSD approach and different basis sets.}
\label{TResult1}
\begin{tabular}{lllll}
    \hline
    \hline
    $m_a$, eV &$\ $ basA $ \ $ & basB $ \ $& basC $ \ $ & basD (Final)\\
    \hline
    $10  $  & $+3.26 \times 10^{-5}$ 
            & $+3.32 \times 10^{-5}$ 
            & $+3.32 \times 10^{-5}$
            & $+3.36 \times 10^{-5}$\\

    $10^2$  & $+3.22 \times 10^{-5}$ 
            & $+3.32 \times 10^{-5}$
            & $+3.30 \times 10^{-5}$ 
            & $+3.36 \times 10^{-5}$\\

    $10^3$  & $+2.76 \times 10^{-5}$ 
            & $+2.86 \times 10^{-5}$ 
            & $+2.84 \times 10^{-5}$ 
            & $+2.90 \times 10^{-5}$ \\

    $10^4$  & $+3.50 \times 10^{-6}$ 
            & $+3.84 \times 10^{-6}$
            & $+3.84 \times 10^{-6}$ 
            & $+3.98 \times 10^{-6}$ \\

    $10^5$  & $-1.78 \times 10^{-5}$
            & $-1.83 \times 10^{-5}$
            & $-1.83 \times 10^{-5}$ 
            & $-1.85 \times 10^{-5}$ \\

    $10^6$  & $-1.32 \times 10^{-5}$
            & $-1.35 \times 10^{-5}$
            & $-1.35 \times 10^{-5}$ 
            & $-1.36 \times 10^{-5}$ \\

    $10^7$  & $-5.58 \times 10^{-6}$
            & $-5.74 \times 10^{-6}$
            & $-5.74 \times 10^{-6}$
            & $-5.80 \times 10^{-6}$\\

    $10^8$  & $-9.30 \times 10^{-9}$
            & $-1.02 \times 10^{-8}$ 
            & $-1.02 \times 10^{-8}$
            & $-1.04 \times 10^{-8}$\\

    $10^9$ & $-9.83 \times 10^{-11}$ 
           & $-1.11 \times 10^{-10}$  
           & $-1.11 \times 10^{-10}$
           & $-1.15 \times 10^{-10}$\\

    $10^{10}$ & $-9.85 \times 10^{-13} $  
              & $-1.11 \times 10^{-12}$ 
              & $-1.11 \times 10^{-12}$
              & $-1.15 \times 10^{-12}$ \\

    \hline 
    \hline
\end{tabular}
\end{table}

As one can see from Tables \ref{TResult0} and \ref{TResult1}, results for low-mass axionlike particles are weakly dependent on the basis set size. However, according to our findings, basis functions with large exponents localized on the heavy atom 
should be included in the basis set to describe correctly the electronic wave function asymptotic in the vicinity of a heavy nucleus in the heavy axion case. 
Therefore, the results obtained within the basD basis set are used for  further estimations below.
It can be seen from comparison of Tables~\ref{TResult0} and~\ref{TResult1} that the role of correlation effects increases for high $m_a$ values 
\footnote{In the case of the large axion mass the mechanism of the correlations contribution is the following.  The effect is mainly due to  the valence wave function of the unpaired electron. For large axion masses the effect is determined by the behavior of this wave function in the region close to the nucleus. Correlations increase binding energy and decrease radius of the unpaired orbital, the decrease of the orbital size  leads to the increase of the amplitude of this wave function near the nucleus and thus can increase the effect.}.
This trend can be compared with the contribution of correlation effects to the other properties. For example, in the case of the RaF molecule, which formally has a similar electronic configuration of the heavy atom, correlation effects change the value of the molecule frame dipole moment by 19\%~\cite{Petrov:2020} and the value of the hyperfine structure constant induced by a heavy nucleus by about 32\%~\cite{Skripnikov:2020e}. The former property is mainly determined by the valence region of the valence wavefunction, while the latter one is determined by the core region of the valence wavefunction.

In the region from $10^4 \ \textrm{eV}$ to $10^5 \ \textrm{eV}$ the $\WaxeN$ constant changes its sign. This can be explained by the fact that the effects for low-mass and high-mass axions arise from different distances~\cite{stadnik:2018}. 
The provided results also reproduce the other trends, obtained in Ref. \cite{stadnik:2018}: $\WaxeN$ is almost independent of $m_a$ for low-mass axions and is determined solely by electronic shell properties; for high-mass axions $\WaxeN$ is scaled as $m_a^{-2}$
\footnote{Note, that results of Ref. \cite{stadnik:2018} are provided there with the factor proportional to $m_a^2$, while we present $\WaxeN$ values without this factor.}.
Both these properties can be explained by the Yukawa-type potential properties: for light axions the  Yukawa range is significantly bigger than the size of a molecule and the exponential factor plays no role, and in the opposite case of $m_a \rightarrow +\infty$ the asymptotic behavior takes place: 
\begin{equation}
\frac{e^{-m_a r}}{4 \pi r} 
\sim
\frac{1}{m_a^2} \delta (\mathbf{r}).
\end{equation}
Some analysis can also be found in Supplemental Material of Ref. \cite{stadnik:2018}.
Below we discuss separately the low-mass and the high-mass axion cases. In the latter case, the potential (\ref{Yukawa}) is spatially localized in the vicinity of the nucleus and in this case the $\WaxeN$ parameter belongs to the class of the ``atom in a compound'' characteristics \cite{Titov:14a,Skripnikov:15b,Skripnikov:2020e}. This means that it should have similar correlation trends to other characteristics such as the effective electric field acting on the electron electric dipole moment, etc.

\textit{Low-mass limit.} As one should expect from the form of the potential (\ref{Yukawa}) and as it can be seen from Table \ref{TResult1}, for $m_a \ll 10^3~\textrm{eV}$ the $\WaxeN$ parameter is almost independent of $m_a$.
Indeed, since for low-mass axions $\WaxeN$ constant does not depend on the axion mass,  $\WaxeN$ is determined mainly by the radius of valence electrons orbitals  which carry electron angular momentum. Therefore,  axion mass is replaced by  $\sim \hbar/(c a_B)$  ($a_B$ is the Bohr radius).
Thus, for the low-mass axion case, we can explicitly obtain the link between the $\bar{g}_N^s g_e^p$ product and the energy shift $\delta E$ defined by \Eref{Etp} employing the electronic structure calculation. The mass of the axionlike particle becomes unimportant here. The current limits for this interaction constants product $|\bar{g}_N^s g_e^p|/(\hbar c) \lesssim  10^{-19} $~\cite{stadnik:2018} can be obtained from the interpretation of the  second generation of the ThO experiment  \footnote{Note, that the limit $|g_N^s g_e^p|/(\hbar c) \le 1.3 \times 10^{-18}$ in Ref. \cite{stadnik:2018} was obtained from the HfF$^+$ experiment \cite{Cornell:2017}; it is stronger than the limit that follows from the first generation of the ACME experiment \cite{ThO,stadnik:2018}, but not as strong, as the consequence of the second generation~\cite{ACME:18}.}.
Note, that this constraint is the strongest for $m_a \gtrsim 10^{-2}~\textrm{eV}$ (see Fig.~2 in Ref.~\cite{stadnik:2018}).
Substituting the final value 
$$\WaxeN (m_a \ll 1 \textrm{keV}) \simeq 3.36 \times 10^{-5} \frac{m_e c }{\hbar}$$
and the value of the mentioned constraint on $|\bar{g}_N^s g_e^p|$ into \Eref{Etp}, one can obtain the expected upper limit of the energy shift in the YbOH molecule $\delta E \simeq 200\ \mu\textrm{Hz}$. This value
is only slightly lower than the sensitivity to $T,P$-violating effects in the most sensitive current experiments ~\cite{ACME:18,Maison:20a}. 
In Ref.~\cite{kozyryev2017precision} it was suggested that by using the YbOH molecule, the sensitivity to the electron EDM and other parameters of $T,P$-violating interactions could be increased significantly.
Therefore, it is expected that the sensitivity of the YbOH experiment 
may be enough to set stronger constraints on $|\bar{g}_N^s g_e^p|$.

\textit{High-mass limit.} For axions with mass $m_a\gg~1~\textrm{MeV}$, the approximate dependence is $\WaxeN(m_a) \simeq \widetilde{W} m_a^{-2}$, where the $\widetilde{W}$ value does not depend on $m_a$. 
This allows one to parametrize the energy shift as
\begin{equation} 
    \delta E \approx  \frac{\bar{g}_N^s g_e^p}{m_a^2} \Omega\widetilde{W},
\end{equation}
where
$$\widetilde{W} = \lim_{m_a \rightarrow +\infty} m_a^2  \WaxeN(m_a).$$
According to \tref{TResult1}, 
$$|\widetilde{W}| \approx 1.15 \times 10^{-10} \ \textrm{GeV}^{2}\times \frac{m_e c }{\hbar}.$$ 
The order of magnitude constraint $|\bar{g}_N^s g_e^p|/(\hbar c m_a^{2}) \lesssim 10^{-14}\  \textrm{GeV}^{-2}$ can be obtained from the interpretation of the experiment with the ThO molecule \cite{stadnik:2018,ACME:18}. The energy shift in the YbOH molecule corresponding to this constraint would be $\sim$70~$\mu$Hz. 
This value is only few times lower than 
the effect in the ThO molecule~\cite{stadnik:2018}.
As it was noted above, the expected sensitivity of the YbOH experiment should be high enough to set updated constraints.

\section{Conclusion}
The effect of the electron-nucleus interaction mediated by axionlike particles in the YbOH molecule is studied. For this, we have calculated the molecular constant that characterizes this interaction within the relativistic coupled cluster theory and studied contributions of correlation effects. According to the obtained results, one can conclude that the planned experiment to search for the permanent molecular $T,P$-violating EDM can also be sensitive to the considered axion-mediated effect and is expected to set updated strong constraints on the corresponding coupling constants depending on the mass of the axionlike particles. The explicit dependence on the axion mass is studied in the limiting cases of extremely light and extremely heavy particles; the link of the corresponding energy shift with the coupling constants is reported.

\begin{acknowledgments} 
Electronic structure calculations have been carried out using computing resources of the federal collective usage center Complex for Simulation and Data Processing for
Mega-science Facilities at National Research Centre “Kurchatov Institute”, http://ckp.nrcki.ru/.

$~~~$Molecular coupled cluster electronic structure calculations have been supported by the Russian Science Foundation Grant No. 19-72-10019. Calculations of the $\WaxeN$  matrix elements were supported by the foundation for the advancement of theoretical physics and mathematics ``BASIS'' grant according to Projects No. 20-1-5-76-1 and No. 18-1-3-55-1. Calculation of the Gaunt contribution has been supported by Russian Foundation for Basic Research Grant No. 20-32-70177. V.F. acknowledges  support by the Australian Research Council Grants No.  DP190100974 and No. DP20010015.  N. R. H. acknowledges support by the Gordon and Betty Moore Foundation (Grant No. 7947) and the Alfred P. Sloan Foundation (Grant No. G-2019-12502).
\end{acknowledgments} 

\bibliographystyle{apsrev}

\end{document}